\documentclass{ws-procs9x6} 
\usepackage{lineno}
\usepackage{hyperref}

\begin{document}
\title{Central exclusive meson production in \\
  proton$-$proton collisions in ALICE at the LHC}

\author{R. Schicker \\
(for the ALICE Collaboration)}

\address{Physikalisches Institut\\
University Heidelberg  \\
69120 Heidelberg, Germany \\
E-mail: schicker@physi.uni-heidelberg.de}

\begin{abstract}
Central exclusive production at hadron colliders is characterised  by the
hadronic state produced at or close to midrapidity, and by the two forward
scattered protons, or remnants thereof. No particles are produced between
the midrapidity system and the forward going beam particles, and such
events can hence be identified experimentally by a double-gap topology.
At LHC energies, central exclusive production in proton$-$proton collisions
is dominated by pomeron$-$pomeron fusion. The models to describe such reactions
are reviewed, and the ongoing efforts in the ALICE Collaboration to analyse
double-gap events taken in Run 2 at the LHC are presented.
\end{abstract}


\bodymatter

\vspace{-.04cm}
\section{Diffractive event topologies}\label{sec:sec1}

In the Regge approach, the hadronic interaction between strongly interacting
particles is due to the exchange of trajectories. Such trajectories
$\alpha (t)$ parameterise the almost linear dependence between the spin and
the mass squared ($t\!=\!m^{2}$) of a particle and its higher spin excitations.
A trajectory is characterised by two parameters, the intercept
$1\!+\!\varepsilon$ and the slope $\alpha'$, and is expressed as
$\alpha(t) = 1\!+\!\varepsilon\!+\!\alpha' t$. In Regge phenomenology, the
contribution of a trajectory to the energy behaviour of the total cross
section scales as $\sigma_{tot} \propto s^{\varepsilon}$. For meson trajectories,
such as the $\rho,\omega,a,f$, the value of $\varepsilon\!\sim\!-0.5$ and the
corresponding contributions decrease as function of energy \cite{PomQCD}.
Such a behaviour is seen in the elastic and total hadron-hadron cross section
which decrease from threshold up to a centre-of-mass energy
\mbox{$\sqrt{s} \sim$ 20 GeV.} At higher energies, however, elastic and total
hadron-hadron cross sections increase \cite{PDG}. Within the Regge approach,
this increase is attributed to the existence of an additional exchange
mechanism in both elastic and inelastic channels, which is refered to as
the pomeron trajectory. The characterisation of the bound states underlying
the pomeron trajectory, and their possible experimental verification,
is a considerable challenge from the theoretical as well as the
experimental perspective.

\begin{figure}
\begin{center}
  \vspace{-.6cm}
  \includegraphics[width=.27\textwidth]{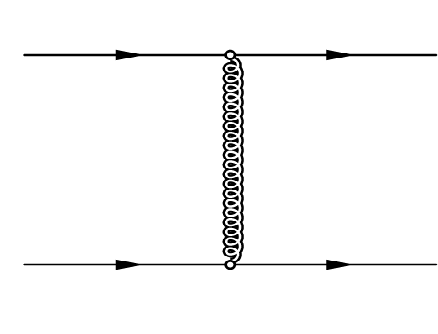}
  \includegraphics[width=.28\textwidth]{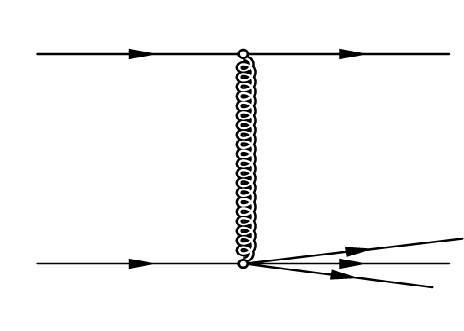}
  \includegraphics[width=.28\textwidth]{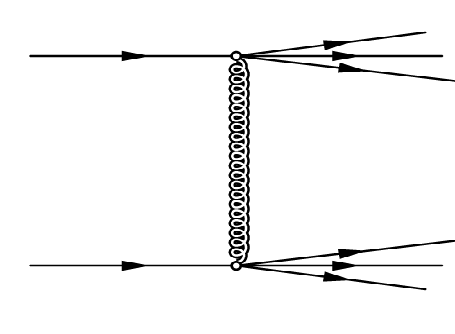}
  \label{fig:topo1}
\end{center}
\end{figure}
\begin{figure}
  \begin{center}
\vspace{-1.6cm}
  \includegraphics[width=.28\textwidth]{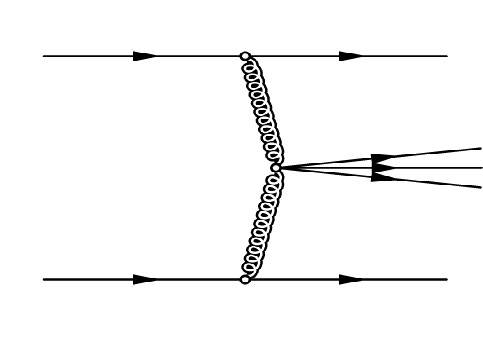}
  \includegraphics[width=.28\textwidth]{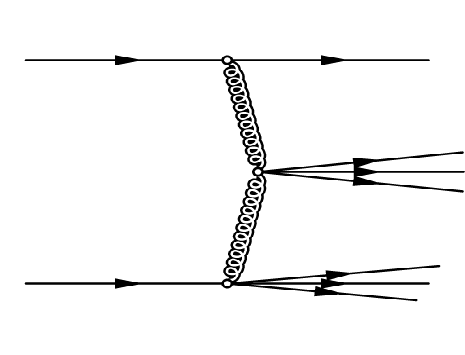}
  \includegraphics[width=.28\textwidth]{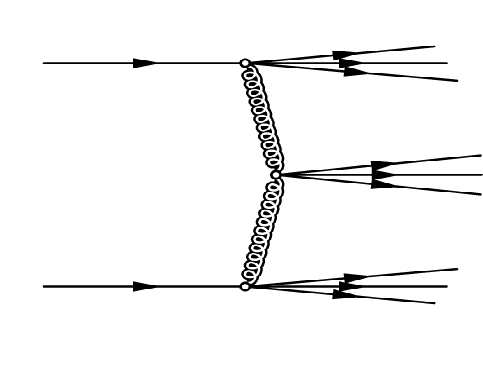}
  \vspace{-.4cm}
  \caption{Top row: Single pomeron elastic scattering (left), and single
    pomeron inelastic scattering (middle and right).
Bottom row: Central exclusive production (left), central production with
proton excitation (middle and right).}
    \label{fig:topo}
\end{center}
\end{figure}
\vspace{-.4cm}
In the top row of Fig. \ref{fig:topo}, the diagrams of strong elastic
and inelastic scattering due to single pomeron exchange are shown. The
corresponding cross sections have been measured at LHC energies by the
TOTEM and the ALICE Collaboration, respectively \cite{TOTEM,ALICE1}.
In the bottom row of Fig. \ref{fig:topo}, the double-pomeron exchange
channel of exclusive production is shown on the left, and central
production with proton excitation in the middle and on the right.
The universality of pomeron exchange can be tested in these central
production channels at high energies, where Regge meson contributions are
negligible as compared to the pomeron exchange.
\vspace{-.3cm}
\section{The spin structure of the soft pomeron}\label{sec:sec2}

It is generally agreed that the pomeron exchange carries the vacuum internal
quantum numbers of charge $Q=0$, colour charge $Q_{C}=0$, isospin $I=0$ and
charge conjugation $C=1$. Under intense debate is, however, the spin structure
of the pomeron exchange in the nonperturbative regime of QCD, the soft pomeron
exchange \cite{softPom}. In contrast to the vacuum expectation value of spin
zero, a vector pomeron exchange has been successfully used to describe elastic
proton$-$proton and proton$-$antiproton scattering data, as well as production
of high-$p_{\rm T}$ jets and Drell-Yan pairs \cite{DL1,DL2}. In this vector
approach, the single pomeron exchange amplitudes of proton$-$proton and
proton$-$antiproton elastic scattering carry, however, a relative minus sign.
By use of the optical theorem, this minus sign propagates into an opposite
sign between the proton$-$proton and proton$-$antiproton cross section.

Alternatively, a description of the soft pomeron as an effective \mbox{rank-2}
symmetric tensor has been proposed \cite{Lebie1}. Here, the meson exchanges
of positive and negative charge conjugation are treated as effective tensor
and vector exchanges, respectively, with many effective couplings to hadrons
derived from experimental data. The single spin asymmetry data of the STAR
Collaboration have been analysed with the tensor pomeron
model \cite{STAR,Ewerz1}. The results of this analysis disfavour the scalar
pomeron, and are in very good agreement with the tensor pomeron. 
In the analysis of central production of scalar and pseudo-scalar mesons
within the vector and tensor pomeron approach, the azimuthal opening angle
between the two final-state protons was found to be
a sensitive probe for the two approaches \cite{Lebie1}.

\section{DAMA model of central production}\label{sec:sec3}

The double-differential cross section d$\sigma$/d$M$d$p_{\rm T}$ for
central exclusive meson production shown in the bottom left of
Fig. \ref{fig:topo} can be derived in a model based on a Dual Amplitude
with Mandelstam Analyticity (DAMA) \cite{Bugrij}. For central production,
the direct-channel pole decomposition of the DAMA amplitude is
relevant \cite{FJS1}. In this approach, the pomeron flux in the
proton is convoluted with the pomeron-pomeron-meson cross section
derived by the optical theorem from the DAMA amplitude \cite{FJS2}.
The DAMA amplitude is based on nonlinear, complex trajectories.
The imaginary part of the trajectory is related to the decay width
of the bound states of the trajectory, and is connected to the
almost linear real part by a dispersion relation.   

\begin{figure}
\begin{center}
\vspace{-.5cm}
\includegraphics[width=.49\textwidth]{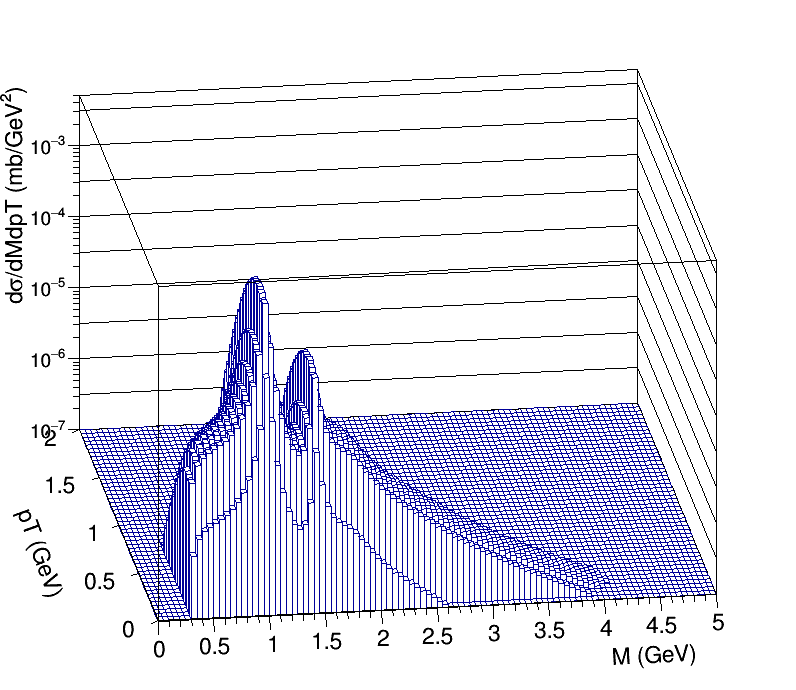}
\includegraphics[width=.49\textwidth]{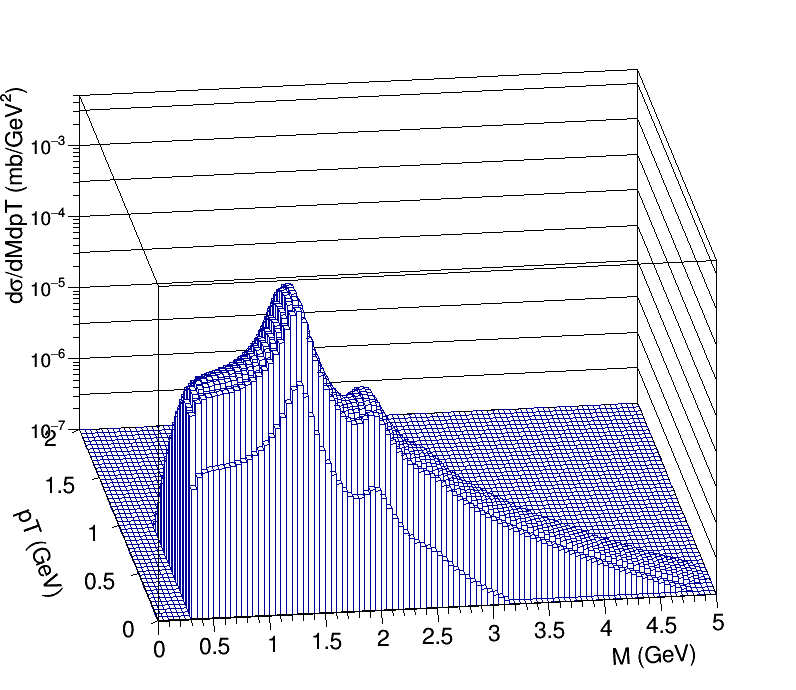}
\caption{Central exclusive cross section for resonances on the  $f_{0}(980)$
  trajectory (left), and on the $f_{2}(1270)$ trajectory (right)
  (Figs. from Ref. \citenum{FJS2}).}
\label{fig:DAMA}
\end{center}
\end{figure}
\vspace{-.4cm}
In Fig. \ref{fig:DAMA} the cross section d$\sigma$/d$M$d$p_{\rm T}$
for the resonances on the trajectory of the $f_{0}(980)$ is shown
on the left, and the $f_{2}(1270)$ on the right.

\section{Central exclusive production in ALICE}\label{sec:sec4}

ALICE consists of a  central barrel in the pseudorapidity range
$|\eta| < 0.9$, and a forward muon spectrometer \cite{ALICEperf}.
An inner tracking system (ITS), composed of silicon pixel, drift
and strip detectors, measures the particle tracks in the innermost
6 layers of the central barrel.
A cylindrical time projection chamber with a sensitive volume
from $r_{\rm min}$=88 cm to $r_{\rm max}$=250 cm records track ionisation
clusters. The specific Bethe-Bloch ionisation loss is used for
particle identification, and the cluster space information is used
together with the ITS space points to reconstruct the track momentum.
Additional detectors outside of the central
barrel are used for trigger purposes, and for classifying the event
according to multiplicity. Each of the V0 and AD detectors consists of
a pair of plastic scintillators which cover the
range \mbox{$2.8\!<\!\eta\!<\!5.1$, $4.9\!<\!\eta\!<\!6.1$,}
respectively, and
\mbox{$-3.7\!<\!\eta\!<\!-1.7$,$-7.0\!<\!\eta\!<\!-4.8$.}

\begin{figure}
  \begin{center}
\vspace{-.4cm}
  \includegraphics[width=.66\textwidth]{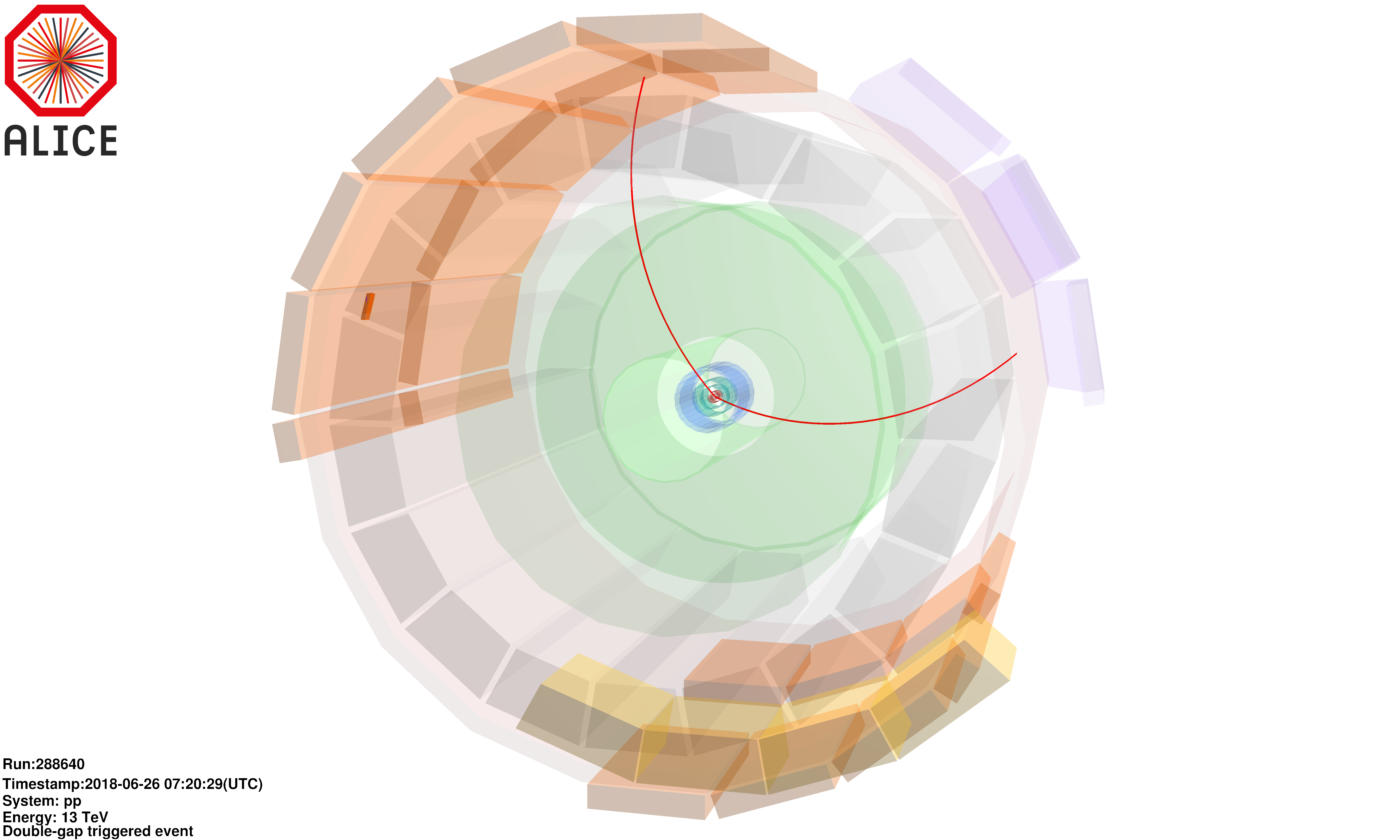}
\caption{Double-gap event in the ALICE central barrel.}
\label{fig:DGevent}
  \end{center}
\end{figure}
\vspace{-.7cm}
The double-pomeron events of Fig. \ref{fig:topo} consist of particles
produced at midrapidity, and of the beam protons, or remnants thereof,
at or close to beam rapidity. No particles are produced in between, which
yields a rapidity region without particles, known as the gap. This rapidity
gap is a signature of central production which can be used for filtering
such events from minimum-bias events in the offline analysis,
or for tagging at the trigger level. A double-gap event in
the \mbox{ALICE central barrel is shown in Fig. \ref{fig:DGevent}.}
\vspace{-.4cm}
\section{ALICE data on exclusive pion pair production}\label{sec:sec5}

\begin{figure}
\begin{center}
\vspace{-0.5cm}  
\includegraphics[width=.31\textwidth]{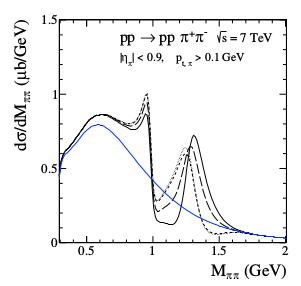}
\includegraphics[width=.32\textwidth]{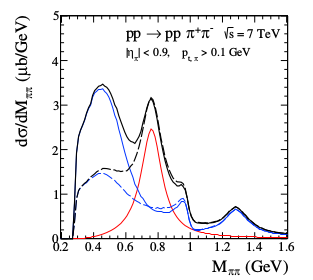}
\includegraphics[width=.35\textwidth]{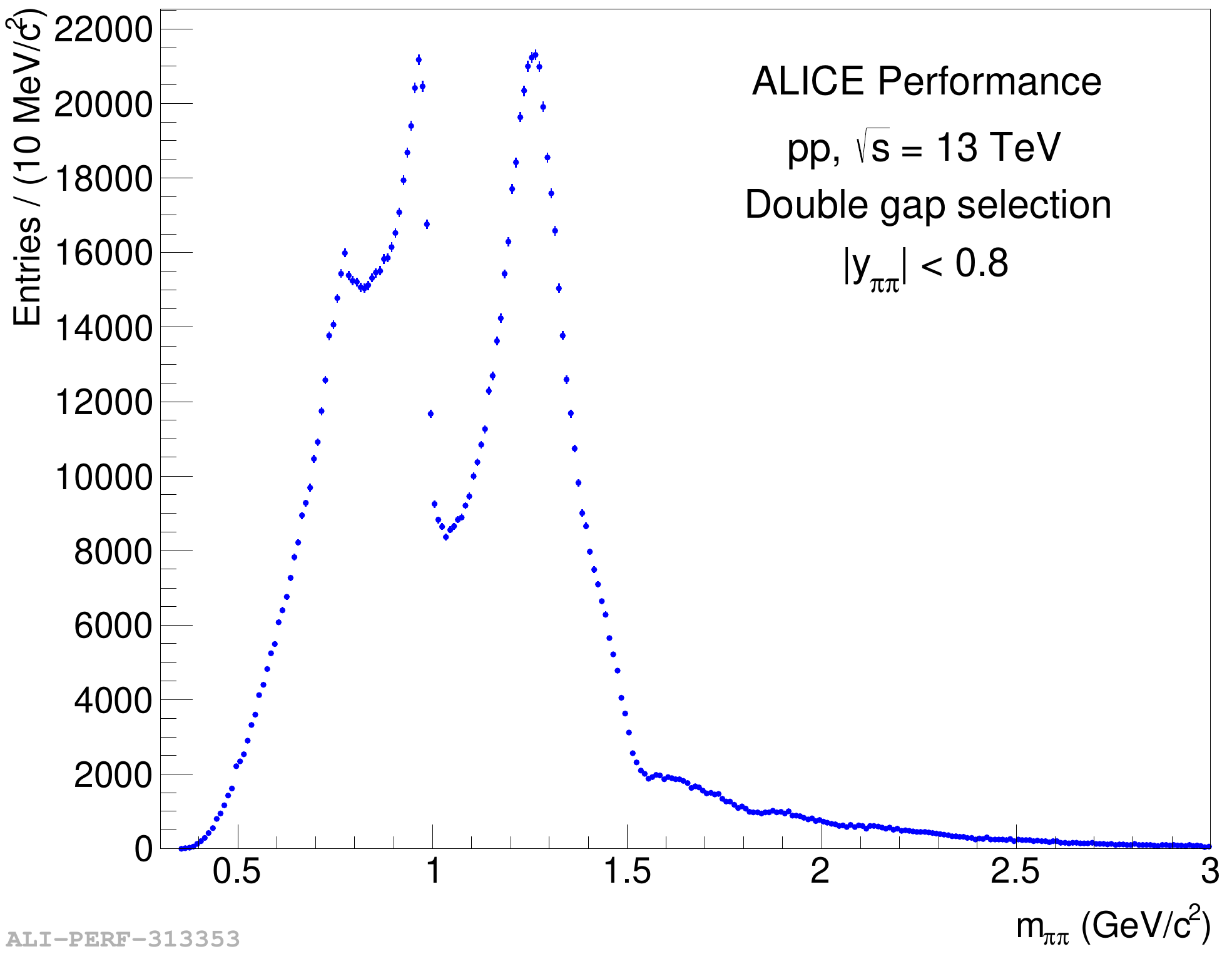}
\vspace{-0.8cm}  
\caption{Model predictions of Ref. \citenum{Lebie2} for central pion
  pair mass distribution (left and middle), and pion pair mass
  distribution measured by ALICE (right).}
\label{fig:pionpairs}
\end{center}
\end{figure}

Exclusive production of $\pi^{+}\pi^{-}$ pairs has been studied
within the tensor pomeron approach \cite{Lebie2}. The dipion
continuum and pion pairs from decays of scalar $f_{0}(500)$,
$f_{0}(980)$ and tensor $f_{2}(1270)$ resonance were considered.
The resulting mass distribution is shown in
Fig. \ref{fig:pionpairs} on the left. Photoproduction of
$\rho(770)$  is visible at LHC energies due to the large
photon fluxes, shown in the middle. The uncorrected pion-pair
mass distribution measured by ALICE in proton$-$proton collisions
at $\sqrt{s}$ = 13 TeV  is shown on the right.  An analogous
analysis has been done for $K^{+}K^{-}$-pairs \cite{Lebie3}.
The ALICE data sample shown in \mbox{Fig. \ref{fig:pionpairs}}
will be increased in the LHC Run 3 by a factor
of \mbox{about 40.} The study of the measured resonances
by a Partial Wave Analysis in both the light quark and
strangeness sector \mbox{will become feasible.}

\vspace{-1.2cm}  
\section{Acknowledgments}
\vspace{-.1cm}  
This work is partially supported by the German Federal Ministry of Education and
Research under promotional reference 05P19VHCA1.

\end{document}